%% file: scientistsmigration.tex
\documentclass[nologo,11pt,a4paper]{ETHpaper}

\usepackage{graphicx}
\graphicspath{{./figures/}}
\usepackage[compress]{natbib}
\usepackage{amsfonts}
\usepackage{array}
\usepackage{booktabs}
\usepackage{amsmath}
\usepackage{caption}

\usepackage{color}
\usepackage{multirow}
\usepackage{soul}
\newcommand{\tnode}[2]{(#1 {\textrm{-}} #2)}
\newcommand{\Tb}[1]{\mathbf{T}^{(#1)}}
\newcommand{\Tbt}[1]{\mathbf{\tilde{T}}^{(#1)}}

\usepackage{lmodern}

\begin{document}

\title{The Mobility Network of Scientists: \\Analyzing Temporal Correlations in Scientific Careers}
\titlealternative{The Mobility Network of Scientists. (Submitted, 28.02.2020)}

\author{Giacomo Vaccario$^{1\ast}$, Luca Verginer$^{1}$, Frank Schweitzer$^{1}$}
\authoralternative{G. Vaccario, L. Verginer, F. Schweitzer}
\address{\bigskip \small
$^1$ETH Z\"urich, Chair of Systems Design, Department of Management, Technology and \\ Economics, Weinbergstrasse 56/58, CH-8092 Z\"urich, Switzerland

\bigskip
$^\ast$Corresponding author; E-mail: gvaccario@ethz.ch
}

\maketitle

\begin{abstract}
  The mobility of scientists between different universities and countries is important to foster knowledge exchange.
  At the same time, the potential mobility is restricted by geographic and institutional constraints, which leads to temporal correlations in the career trajectories of scientists.  
  To quantify this effect, we extract 3.5 million career trajectories of scientists from two large scale bibliographic data sets and analyze them applying a novel method of higher-order networks.  
We study the effect of temporal correlations at three different levels of aggregation, universities, cities and countries.
We find strong evidence for such correlations for  the top 100 universities, i.e. scientists move likely between specific institutions.
These correlations also exist at the level of countries, but cannot be found for cities. 
Our results allow to draw conclusions about the institutional path dependence of scientific careers and the efficiency of mobility programs.

\end{abstract}

\section{Introduction}
\label{sec:intro}
Mobility of high-skill labour is a hot topic in economics, politics and research policy.
One of the most prominent manifestations of this phenomenon is the international mobility of scientists.
Global mobility of scientist has increased over the past decade and is regarded by the \citet{OECD2017} as being a ``key driver of knowledge circulation worldwide''.
In the past this phenomenon has caused fears of ``Brain Drain'', whereby high-skilled labour moves abroad to the benefit and detriment of the receiving and sending country, respectively.
This pessimistic view has been challenged by recent research showing that both sending and receiving countries may benefit from this labour mobility~\citep{Saxenian2005, Agrawal2011, Petersen2018}.
Nevertheless, modern economies rely on high-skill labour to maintain their competitive advantage~\citep{Chambers1998, Beine2001, Beechler2009, Bahar2012}.
For this reason, attracting and retaining scientists are becoming a key concern for migration policy~\citep{Boucher2014}.

Following the arguments from above, scientist mobility plays a central role in the exchange of knowledge.
This implicitly assumes that scientists can move relatively freely and that there is a high degree of ``mixing''.
Mixing in this context means that different scientific careers are possible, and they do not always follow the same pattern.
However, we know that geography constrains scientists' career trajectories~\citep{verginer2018brain} in particular, and research and development activities in general~\citep{scholl2018spatial}.
Additionally, \citet{Deville2014,clauset2015systematic} have shown that institutional prestige constrains scientific careers.
Indeed, both geography and prestige are essential to capture complex features of scientists' mobility~\citep{vaccario2018mobility}.
In this work, we test if scientific careers are indeed global and free, resulting in a worldwide scientist mobility network with high levels of exchange and mixing.
Or alternatively, if scientific careers often follow similar patterns and thus lead to weak mixing.
We test this hypothesis by looking at scientific careers among the 100 top universities as ranked by the Times Higher Education \cite{THEurl}.

\begin{figure}
  \footnotesize
  \includegraphics[width=0.97\textwidth]{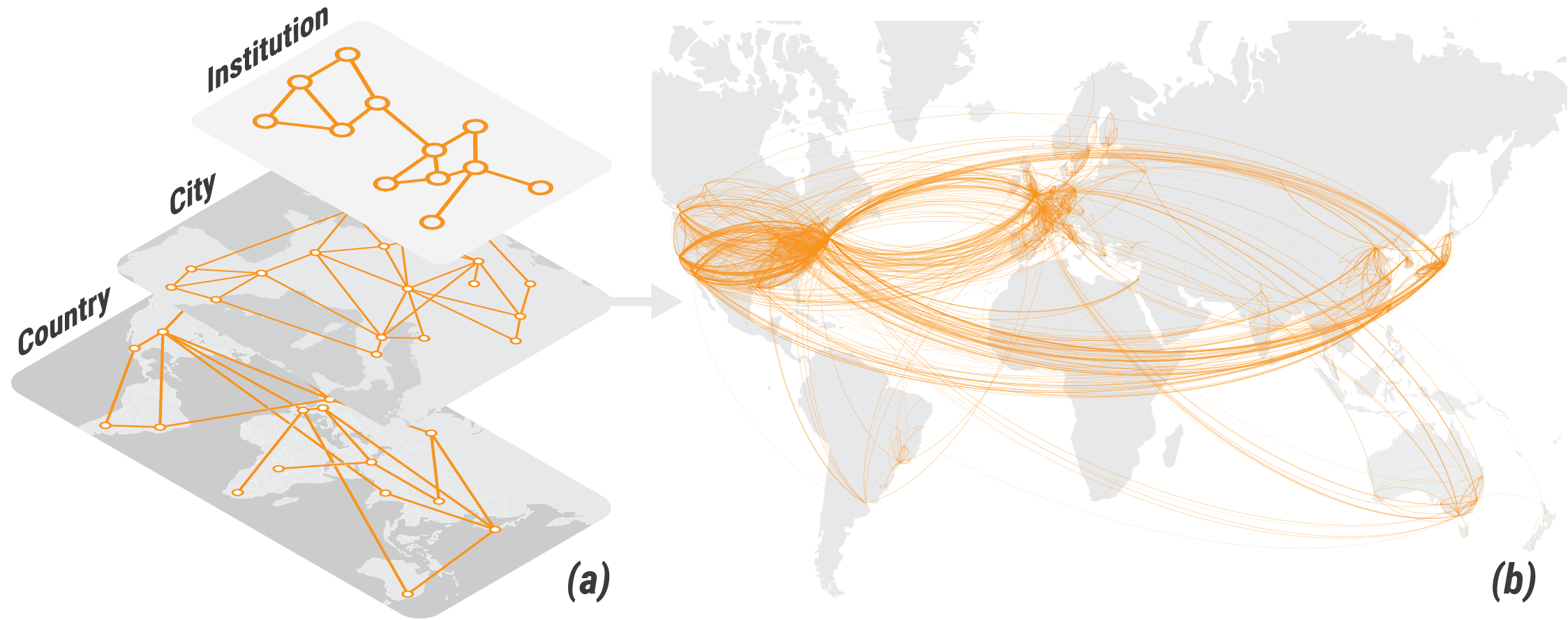}
  \caption{The affiliation data can be aggregated from affiliation level (e.g., university) to city and country as shown in panel (a). In panel (b) the top 10\% most common paths at city level in 2004 for the MEDLINE corpus are shown.}\label{fig:multilevel}
\end{figure}
Previous works have projected career trajectories at the city, regional and country level to study their patterns and economic impact~\citep{Miguelez2014, Petersen2018,verginer2018brain}.
These projections are obtained by aggregating trajectories traversing institutions located in the same cities, regions or countries.
By doing so, scientists' mobility can be studied at different levels of aggregation: affiliation, city and country levels (see Fig.~\ref{fig:multilevel}).
For this reason, we complement our analysis by looking at career trajectories beyond the 100 top universities and consider also the city and
country level.
Hence with our study, we explore academic mobility at three different levels of aggregation.

To empirically address the importance of academic mobility and consequences for knowledge exchange, we reconstruct the career trajectories of scientists through bibliographic data.
Specifically, we use the affiliations reported on scientific publications, to track scientific careers.
We refer to these time-ordered sequences institutions of individual scientists as the ``career trajectories'', in line with the literature on labour mobility.
Previous research~\citep{clauset2015systematic} has aggregated these trajectories to construct a static network where nodes represent research institutions and link the flow of scientists between them.
However, recent advances in computer and network science have raised concerns over the naive aggregation of temporal sequences into static networks.
In particular, neglecting temporal correlations can lead to erroneous conclusions about the accessibility and importance of nodes~\citep{lentz2013unfolding} and the possible dynamic processes unfolding on networks~\citep{pfitzner2013betweenness}.
To address this issue, we use \emph{Higher-Order Networks}, a novel method to represent and model sequential data~\citep{rosvall2014memory,scholtes2014causality,xu2016representing}.
This network representation allows us to capture temporal correlations by means of topological characteristics.
Additionally, \citet{scholtes2017kdd} has combined different higher-order networks to obtain a \emph{Multi-Order Graphical Model} that encode temporal correlations of varying lengths.
We build on these recent developments in network science to contribute to the literature on scientists' mobility in particular and high-skill labour in general.

The rest of the paper is structured as follows.
First, in Sect.~\ref{sec:recon-mob-path}, we introduce the data we will be using for the analysis.
We then describe, in Sect.~\ref{sec:anal-scien-mob}, how career trajectories may be represented using a network perspective and introduce higher-order network models.
In Sect.~\ref{sec:quant-temp-corr}, we estimate the importance of temporal correlations in our data by using multi-order network models.
Moreover, we analyze the implications of the identified temporal correlations for knowledge exchange in Sect~\ref{sec:career-traj-and-mix}.
To complement the analysis at the university level, we also test for the existence of temporal correlations at city and country level, in Sect.~\ref{sec:city-and-country-lev}.
Finally, we discuss our findings and their implications in Sect.~\ref{sec:discussion}.

\section{Reconstructing Mobility Paths}
\label{sec:recon-mob-path}
To study the mobility patterns of scientists, we need a large scale dataset on the institutions, scientists have worked.
We extract this data from publications records.
Indeed, the affiliation is a piece of meta-data available with virtually every published paper.
We rely on this meta-data to reconstruct career trajectories.
Precisely, we extract from two large bibliographic datasets the sequence of affiliations for a given disambiguated author, namely MEDLINE and MAG.
Specifically for the analysis at the institution level, i.e., the 100 most ``prestigious'' universities we use the Microsoft Academic Graph (MAG) dataset, and for the comparison at city and country level, we use the MEDLINE corpus.
Extracting affiliation sequences from these corpora yields sequences of time-stamped (i.e., publication date) locations.
An example of such a record, as found in the MEDLINE corpus, is shown in Table~\ref{tab:example_medline_record}.
This record is equivalent to the records extracted from MAG (see Table~\ref{tab:example_mag_record}), except for the fact that MAG lists the affiliation but not the city or country.
\begin{table}[htb]
  \centering
  \caption[Example Affiliation Record]{Example of career path of a specific author (LM Shul.).
  Only a subset of her publications is shown. For each record we have the year of publication, the city of the affiliation and the relative PubMed ID identifying the paper. 
  }
  \label{tab:example_medline_record}
  \begin{tabular}{ccc}
  \hline
  Year &           City &     Pubmed ID \\
  \hline
  $\vdots$ &      $\vdots$ &   $\vdots$ \\
  2000 &      Miami, FL, USA &  11054153 \\
  2000 &      Miami, FL, USA &  10928576 \\
  2000 &      Miami, FL, USA &  10714670 \\
  2000 &      Miami, FL, USA &  10634252 \\
  2001 &  Baltimore, MD, USA &  11763581 \\
  2001 &  Baltimore, MD, USA &  11391746 \\
  2002 &  Baltimore, MD, USA &  15177058 \\
  $\vdots$ &  $\vdots$ &  $\vdots$ \\
  \hline
  \end{tabular}
\end{table}
\begin{table}[htb]
  \centering
  \caption{Example of career path of a specific author (I HAJR.) in the MAG.
  Only a subset of her publications is shown. 
  For each record we have the year of publication, the affiliation name and the relative MAG ID identifying the paper.}
  \label{tab:example_mag_record}
  \begin{tabular}{ccc}
  \hline
  Year &           Affiliation &     MAG ID \\
  \hline
  $\vdots$ &      $\vdots$ &   $\vdots$ \\
  2012 &      Simon Fraser University &  822D2439 \\
  2012 &      Simon Fraser University &  8126FAC7 \\
  2012 &      Simon Fraser University &  775402B2 \\
  2013 & Centrum Wiskunde \& Informatica &  80F5235B \\
  2014 &  Brown University &  8126FAC7 \\
  2014 &  Brown University&  7CE98F90 \\
  $\vdots$ &      $\vdots$ &   $\vdots$ \\
  2015 & Stanford University& 7E028B00\\
  $\vdots$ &  $\vdots$ &  $\vdots$ \\
  \hline
  \end{tabular}
  \end{table}

For the analysis at affiliation level, we use the \textit{Microsoft Academic Graph} ({MAG}) released for the KDD Cup competition in 2016.
The KDD cup version of the MAG data contains more than 126 Mill.\ publications~\citep{sinha2015overview}.
Each publication is also endowed with various attributes such as unique ID, publication date, title, journal ID, author ID, and affiliations
From this data, we extracted the career trajectories of scientists at affiliation level by using the affiliations reported on their publications.
For the precise details of the extraction procedure see \citet{verginer2018brain}.
We obtain $235\,935$ scientist career trajectories moving through 100 universities.
These 100 university have been disambiguated manually and the list of matched universities is available upon request from the authors.
We use the MAG for the analysis at university level because it has disambiguated authors and covers a wide array of journals across fields.
This means that especially for US and EU based institutions, we have comprehensive coverage of publications.

However, the MAG has limitations that hinder a reliable analysis of scientists' career beyond the 100 top universities. 
First, the string based matching employed by MAG is imperfect.
For example, the MAG contains different affiliation IDs for the campuses of the University of California (e.g., UCLA, UCI, etc.) and a separate ID for the University of California UC.
Hence, when creating careers trajectories considering all the affiliation IDs, one finds trajectories with fictitious locations, e.g., $UCLA \to UC \to UCLA \to UC$.
Note that this is not a problem when analyzing top institutions as the trajectory $UCLA \to UC \to UCLA \to MIT$ would become $UCLA \to MIT$.
Second, there is no unambiguous definition of what constitutes a research center.
For example, the University of California is composed of ten campuses (e.g., UCLA, UCI, etc.), but UC and its campuses have different IDs that do not reflect this hierarchical structure.
Given these issues, to study also the careers of scientists not affiliated with the top 100 universities, 
we do not use affiliations but their location at city and country level.
For this purpose, we use MEDLINE.

MEDLINE is the largest, publicly available bibliographic database in the life sciences and maintained by the U.S. \ National Library of Medicine (NLM).
It contains over 26 Mill.\ papers published in 5,200 journals and 40 languages, goes back in time until 1966 and is continuously updated.
The corpus covers research in biomedicine and health predominantly.
Moreover, we rely on the high-quality dataset by \citet{Torvik2009,Torvik2015} to aggregate universities to the city and the country level.
With this dataset, we can geo-localize the affiliation strings worldwide.
They main disadvantage of MEDLINE is that it has a lower journal and discipline coverage.

We are aware that reconstructing career trajectories using bibliographic data is not without issues.
First, the disambiguation of authors is never perfect.
However, we rely on two high quality and widely used datasets to address this issue \citep{Torvik2009, Torvik2015, sinha2015overview}.
Second, publications are a delayed signal of presence and not a live signal, as the submission and publication date are different.
This means that the affiliation associated with a scientist in a given year might not be her current place of employment.
At the same time, for the present study, we care primarily about the order of affiliations and not their precise timing.
Third, we can only talk about scientists' \emph{mobility} and not about  \emph{migration}, which would require us to know the nationality of the scientists, which we do not have.
We do have information on the country of first publication, which might coincide but is not guaranteed to be correct.

\section{Analyzing scientists' mobility as a network}
\label{sec:anal-scien-mob}

\begin{figure}
  \centering
  \includegraphics[width=0.95\textwidth]{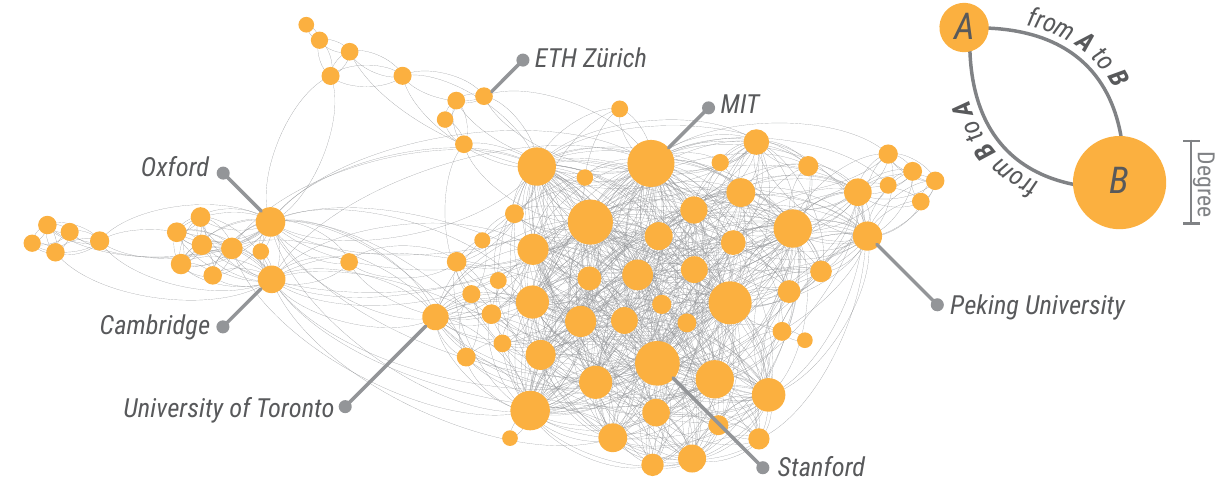}
  \caption{The inter-institution mobility network generated from the 10\% most common paths.
  Nodes represent universities and links scientists moving between them.
  The links are directed and they have to be followed clockwise.
  }\label{fig:first_order_net}
\end{figure}
To study the career trajectories of scientists, we adopt a network perspective.
We represent universities as nodes and scientists' movements between universities as links.
By aggregating the career trajectories, we reconstruct the mobility network that we show in Fig.~\ref{fig:first_order_net}.
From this figure, we see that all universities belong to the same connected component.
This fact implies that there exists a (network) path connecting any pair of universities.
Moreover, the network diameter is six, and the average shortest path is about 2.
Hence, it appears that a scientist from any university could potentially reach any other university within a small number of steps.
In reality, however, both prestige and geography constrain scientists' career trajectories~\citep{clauset2015systematic,verginer2018brain} and thus rule out various potential trajectories. 
There is a ``paradox'' between the topological finding of a short average path and the realized possibilities of such short paths which motivates our more in-depth investigation.

We start our investigation by looking at the temporal correlations in scientists' career trajectories.
Recent studies have shown that mobility patterns with temporal correlations cannot be captured using standard network models~\citep{rosvall2014memory,scholtes2014causality,xu2016representing,scholtes2017kdd}.
For example, these studies have shown that random walk models on networks do not capture travel data of passengers well, due to temporal correlations.
Indeed, most passengers flying from one airport to another take return flights or pass through specific hubs, and hence, the sequences of airports visited exhibit temporal correlations.
To deal with temporal correlations in sequence data, several authors have proposed a new class of network models called \textit{higher-order networks}~\citep{rosvall2014memory,scholtes2014causality,xu2016representing}.

\begin{figure}
  \centering
  \includegraphics[width=0.9\textwidth]{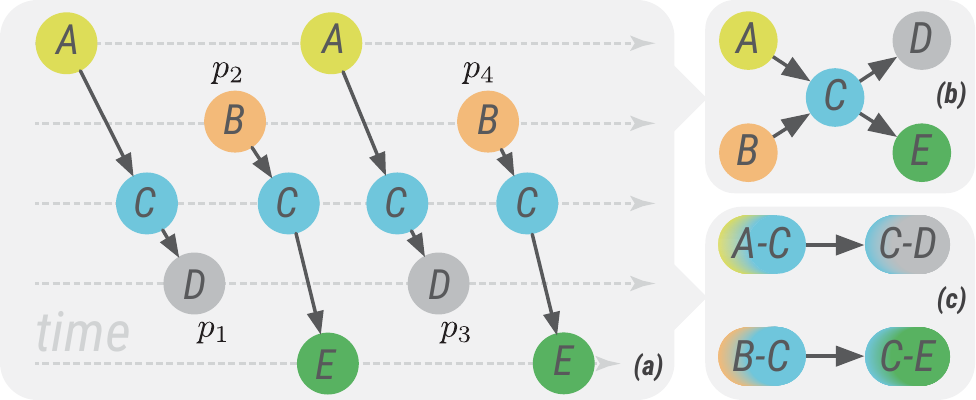}
  \caption{From Temporal sequences to Network representations.
  In panel \textbf{(a)} we have a set of trajectories between the location $A$, $B$, $C$ and $D$.
  Given these trajectories, we can extract several networks.
  The simplest network representation is illustrated in panel \textbf{(b)}, where nodes represent locations and links correspond to the observed moves between these locations. In panel \textbf{(c)}, the same trajectories are encoded in a \textit{second order network} where each node is an observed sequence of two locations.}
  \label{fig:higher_order_illustration}
\end{figure}
Higher-order networks are mathematical objects which retain temporal information normally discarded in standard network models.
For example, we see in Fig.~\ref{fig:higher_order_illustration}~(a) four different career trajectories $p_1=p_2=\{A, C, D\}$ and $p_3=p_4=\{B, C, E\}$.
If we were to represent these as a simple \emph{first-order} network (see Fig.~\ref{fig:higher_order_illustration}~(b)), we would imply that the transitions $A\to D$ and $A\to E$ are equally likely.
However, by doing so, we have discarded the (temporal) information that no trajectories are connecting $A\to E$.
To preserve this information, we represent the trajectories $\{p_1,p_2,p_3,p_4\}$ in a second-order network (see Fig.~\ref{fig:higher_order_illustration}~(c)).
In this network, we have four nodes $\tnode{A}{C}$, $\tnode{C}{D}$, $\tnode{B}{C}$ and $\tnode{C}{E}$ and two links $\tnode{A}{C} \to \tnode{C}{D}$ and $\tnode{B}{C} \to \tnode{C}{E}$. With this second-order network, we now respect the (temporal) order implicit in the data.
Similar to standard networks, a second-order network may be represented using a transition matrix $\Tb{2}$.
In this matrix an element $(\Tb{2})_{ee'}$ is the probability of observing the transition $e'$ (e.g., $C\to D$) after observing the transition $e$ (e.g., $A\to C$).

\begin{figure}
  \centering
  \includegraphics[width=0.78\textwidth]{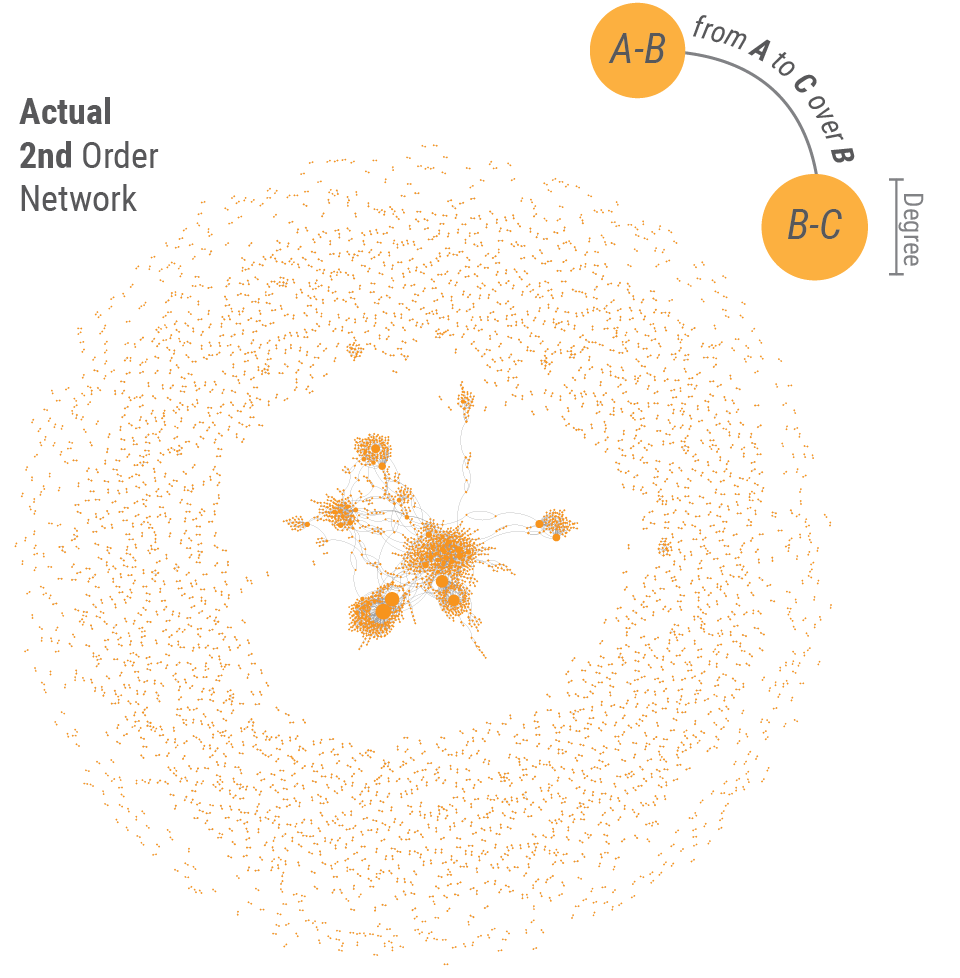}
  \caption{
  The inter-institution mobility network generated from the 10\% most common career trajectories.
  Note that in the second-order network a node represents a first-order edge (e.g., MIT$\to$UCLA) and an edge a link between edges (e.g., (MIT-UCLA)$\to$(UCLA-ETH)).
  The higher-order network represents the \emph{actual} 10\% most common career trajectories.
  }\label{fig:real_memory_illustration}
\end{figure}
\begin{figure}
  \centering
  \includegraphics[width=0.78\textwidth]{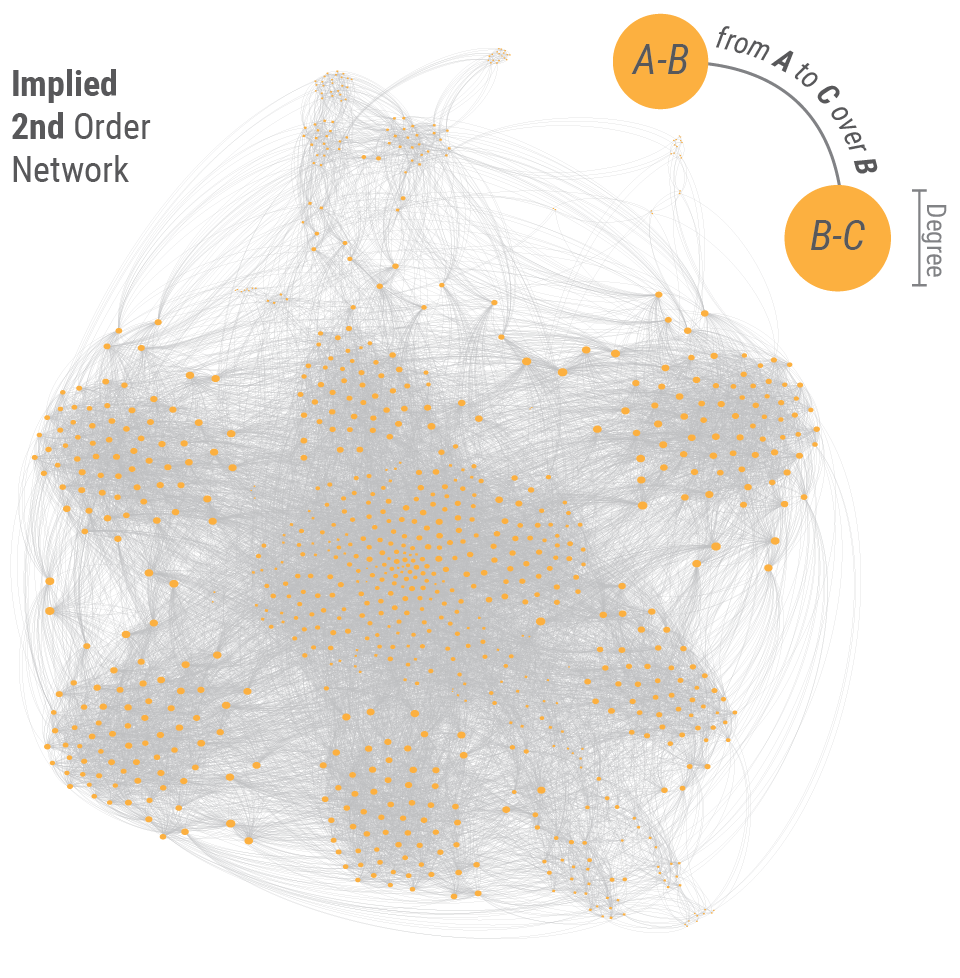}
  \caption{
  The inter-institution mobility network generated from the 10\% most common career trajectories.
  Note that in the second-order network a node represents a first-order edge (e.g., MIT$\to$UCLA) and an edge a link between edges (e.g., (MIT-UCLA)$\to$(UCLA-ETH)).
  The second-order network representation implied by the network shown in Fig.\ref{fig:first_order_net}.
  }\label{fig:implied_memory_illustration}
\end{figure}
In Fig.~\ref{fig:real_memory_illustration}, we represent scientists' career trajectories as a second-order network.
From this visualization, we immediately note two features.
\textit{First}, we note the halo of points that are many disconnected components of size two.
These represent career trajectories that connect only pairs of universities, i.e., trajectories of the type $\tnode{A}{B} \to \tnode{B}{A}$.
\textit{Second}, the largest connected component contains only 23\% of the nodes (i.e., university pairs) and it has a diameter of 24.
These two key statistics indicate that the empirical career trajectories do not connect all university pairs, suggesting the existence of constraints in academic mobility.

We argue that the above finding is caused by the time ordering of universities in career trajectories.
To show this, we depict the second-order network reconstructed using a ``Markovian'' temporal network in Fig.~\ref{fig:implied_memory_illustration}.
This network is constructed from a maximum entropy second-order transition matrix, $\Tbt{2}$~\citep{scholtes2014causality}.
An element $(\Tbt{2})_{\tnode{A}{C},\tnode{C}{E}} = (\Tb{1})_{A,C} \times (\Tb{1})_{C,E}$ where $(\Tb{1})_{A,C}$ is the observed relative frequency of the transition $A \to C$.
By this, we obtain a transition matrix that does not preserve the empirical time ordering of career trajectories.
In Fig.~\ref{fig:implied_memory_illustration}, the network links correspond to all the non-zero entries of the second-order transition matrix of the Markovian temporal network $\Tbt{2}$.
From this network visualization, one would deduce that all universities are connected through career trajectories, and scientists move freely from one university to another.
Indeed, the Markovian second-order network has a diameter of six, and every node belongs to the giant connected component.
The differences between the second-order networks reconstructed using the empirical data and $\Tb{1}$ hint at the presence of temporal correlation in scientists' career trajectories.

\section{Quantifying Temporal Correlations}
\label{sec:quant-temp-corr}
A second-order network captures the temporal correlations of scientists trajectories of length two without loss of information.
In general, trajectories can have different lengths, and hence, can be represented by higher-order networks of different orders.
Then, if we have a sample of career trajectories of different lengths, how do we choose the correct order to model all these trajectories at the same time?
To address this problem, we use the multi-order graphical (MOG) models, and the statistical test developed by~\citep{scholtes2017kdd}.
A MOG-model is a combination of higher-order networks up to order $k$. 
For a MOG-model, it is possible to compute its likelihood given its complexity (i.e., degrees of freedom) and the observed trajectories.
Then, by using a likelihood ratio test between MOG-models with increasing $k$, we choose the model with the \textit{optimal order} $k_{\mathrm{opt}}$.
The MOG-model with a $k_{\mathrm{opt}}$ retains statically significant temporal correlation present in the data without over-fitting.
For details about MOG-model and the test see~\citet{scholtes2017kdd}.

Note that $k_{\mathrm{opt}}=1$ would imply that a first-order network well represents the data, and hence, the trajectories are not significantly influenced by temporal correlations.
In practice, this would mean that the next movement of a scientist depends only on his/her current location.
$k_{\mathrm{opt}}>1$, on the other hand,  would imply that the next location visited by a scientist depends not only his/her current location but also on the previous ones.
To estimate the $k_{\mathrm{opt}}$, We use \texttt{Pathpy}, the open-source path analysis library~\citep{pathpy}. 
When applying the statistical test of \citet{scholtes2017kdd} to scientists' career trajectories, we find $k_{\mathrm{opt}} = 2$.
That means $k_{\mathrm{opt}} > 1$, therefore we have statistical evidence that scientists' mobility exhibits temporal correlations.
Precisely, we find that the next university in a career trajectory depends on a scientist's current and previous universities.

To better understand the source of the identified temporal correlation, we study the temporal motifs of length two.
A temporal motif of length two is a sub-path of length two in a career trajectory, and it can have the following forms: $X\to Y \to X$ (type I) and $X \to Y \to Z$ (type II).
Type I motifs represent pieces of career trajectories where a scientist first changes his/her university ($X \to Y$) and then, goes back to the previous working place ($Y \to X$).
Type II motifs, instead, represent career trajectories of scientists that after changing university ($X \to Y$), do not go back ($Y \to Z$).
For each type of motif, we can compare its empirical probability distribution $P^{(\textrm{\scriptsize emp})}$ to the one expected from a first-order $P^{ (1) }$ and a second-order $P^{ (2) }$  network model.
For more details on how these distributions are computed, see Appendix~\ref{app:motifs}.
If we find that the $P^{(\textrm{\scriptsize emp})}$ of both motif types are similar to $P^{ (2) }$, but not $P^{ (1) }$, we can conclude that the temporal correlations stem from both motif types.
If, on the other hand, only the $P^{(\textrm{\scriptsize emp})}$ of type I (type II) motifs is similar to $P^{ (2) }$,  we can conclude that the temporal correlations stem from type I (type II) motifs.

We start by analyzing the motives of type I, i.e., of the form $X\to Y \to X$.
To compare the $P^{(\textrm{\scriptsize emp})}$, $P^{(1)}$, and $P^{(2)}$ of these motifs, we use the Kullback-Leibler (KL) divergence~\citep{kullback1951information}.
This measure is commonly used to describe the similarity between probability distributions of discrete and unordered variables, and hence, it suits our situation well.

For the motifs of type I, the KL-divergence between $P^{(\textrm{\scriptsize emp})}$ and $P^{(1)}$ is 0.826, while the $P^{(\textrm{\scriptsize emp})}$ and $P^{(2)}$ is 0.038.
The ratio between these two KL-divergences is $\sim 21.7$ and we name it $R_I$.
Hence, the KL-divergence between $P^{(\textrm{\scriptsize emp})}$ and $P^{(2)}$ is more than 20 times smaller compared to the divergence between $P^{(\textrm{\scriptsize emp})}$ and $P^{(1)}$.
This means that the frequency of scientists that go back to their previous working institution is better captured by a second-order network.

For the motifs of type II, the KL-divergence between $P^{(\textrm{\scriptsize emp})}$ and $P^{(1)}$ is 0.798, while the $P^{(\textrm{\scriptsize emp})}$ and $P^{(2)}$ is 0.183.
The ratio between these two KL-divergences is $\sim 4.4$ and we name it $R_II$.
Hence, the KL-divergence between $P^{(\textrm{\scriptsize emp})}$ and $P^{(2)}$ is about 4 times smaller compared to the divergence between $P^{(\textrm{\scriptsize emp})}$ and $P^{(1)}$.
We find again that the second-order model captures motives of type II better.

By comparing $R_I$ and $R_{II}$, we also find  $R_I$ is about five times larger than $R_{II}$.
This result indicates that the temporal correlations in the second-order model are mostly useful to capture motifs of type I.
In the last section of the paper, we discuss and interpret this result.

\section{Career trajectories and mixing}
\label{sec:career-traj-and-mix}
In this section, we analyze the effect that temporal correlations have on \emph{mixing} and diffusion processes unfolding on the scientist mobility network.
We split the analysis of scientist trajectories in a qualitative and quantitative part.
For the former, we compare the alluvial diagrams \citep{lambiotte2019networks} of career trajectories created using
the first- and second-order network models, while for the latter, we compute their entropy growth \citep{scholtes2014causality}.
Note that both the alluvial diagrams and the entropy growth describe and quantify the diffusion process unfolding on a temporal network.
Specifically, for the case of academic mobility, one might think of a diffusion process in which scientists carry knowledge.
Then the question arises, how fast and how far this knowledge would propagate in the network through mobility alone. 
If this happens fast and many institutions are reached, then this corresponds to \emph{strong mixing}.
If on the other hand, few institutions are reached, then we have \emph{weak mixing}.

\begin{figure}[htbp]
  \centering
  \includegraphics[width=0.95\textwidth]{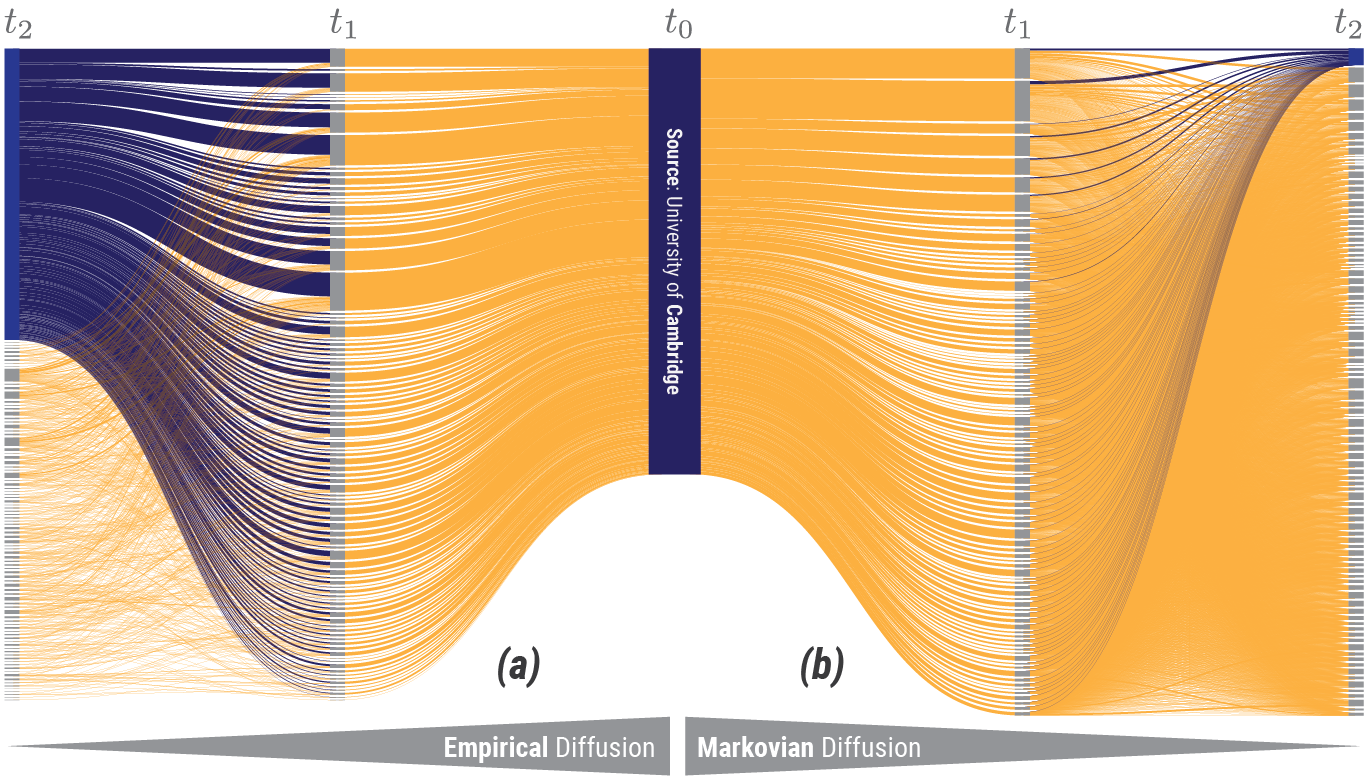}
  \caption{The possible diffusion processes following the career trajectories originating from the University of Cambridge.
  The thickness of an edge represents the probability to observe that move.
  The color of an edge is blue when it represents a move going back to the source node, i.e., the University of Cambridge, otherwise the color is orange.
  The diffusion process describes the probability to find a scientists in a given location after \(t\) moves starting from the source node. On the left side of the plot (a), the empirical diffusion (i.e., actual moves) is shown.
  On the right side of the plot (b), the Markovian diffusion is shown, i.e., the diffusion we would expect if career trajectories had no temporal correlations.
  For visualization purposes, the 50\% most common career trajectories have been used for this figure.
  }\label{fig:alluvial}
\end{figure}

In Fig.~\ref{fig:alluvial}, we show two alluvial diagrams describing the career trajectories originating from the ``University of Cambridge''.
In Fig.~\ref{fig:alluvial}(b), we show the alluvial diagram for the \emph{Markovian diffusion}.
Starting from the University of Cambridge at $t_0$, scientists are expected to move to many institutions with similar probabilities, illustrated by the similar thickens of edges from $t_0$ to $t_1$.
The universities reached at $t_1$ are many, suggesting the existence of strong mixing, and from step $t_1$ to $t_2$, the mixing is amplified.

In Fig.~\ref{fig:alluvial}(a), we show the alluvial diagram for the \emph{empirical diffusion} of career trajectories originating from the University of Cambridge.
In this panel (a), we see that from $t_0$ to $t_1$ many institutions are reached, again suggesting the existence of strong mixing.
However, moving from $t_1$ to $t_2$ this trend is reversed, since most paths originating at the University of Cambridge go back there.
This is visually shown as the blue edges from $t_1$ to $t_2$.
The preferred destination reduces the probability to reach many other locations at $t_2$, implying weak mixing.

Note that the segment from $t_0$ to $t_1$ is identical, albeit ordered differently, for both Fig.~\ref{fig:alluvial}(a) and (b), since both representations capture the first transitions correctly.
The differences in the diffusion process emerge in the second step from $t_1$ to $t_2$, when it matters whether a network is $k_{\mathrm{opt}}=1$ or $k_{\mathrm{opt}}>1$.
As the comparison of Fig.~\ref{fig:alluvial}(a) and (b) demonstrates, disregarding temporal correlations in the Markovian model leads us to overestimate mixing.
Precisely, in Fig.~\ref{fig:alluvial}(a) the returning rate is high and low in Fig.~\ref{fig:alluvial}(b).
This qualitative result is in line with the KL analysis, highlighting the importance of motives of type I, ($X\to Y\to X$).

To quantify the expected mixing of the two models given the observed data, we use the entropy growth ratio introduced by \citep{scholtes2014causality}.
The interpretation of this measure is as follows.
Given an observed transition $e = A\to B$, we compute the corresponding
entropy from the second-order transition matrix, i.e., the probabilities to
transition to the next nodes $e'$:
\begin{equation}
  h_e(\Tb{2}) = - \sum_{e'\in E} (\Tb{2})_{ee'} \log (\Tb{2})_{ee'}
\end{equation}
where $E$ is the set of edges, i.e., possible career moves from one
university to another.

The entropy $h_e(\Tb{2})$ quantifies to what extent the next transition
in a trajectory is determined by the previous transition $e$.
For example, if a row $(\Tb{2})_{e,\cdot}$ has only one element equal to
1, then the transition is deterministic, and hence,  $h_e(\Tb{2}) = 0$
implying weak mixing.
While if all elements of a row are equal, then the next transitions
happen uniformly at random, and hence, $h_e(\Tb{2})$ is high, implying
strong mixing.

To compute the expected overall mixing, one could argue that the total entropy of the model is the sum of all the transition entropies, i.e., $\sum_{e\in E} h_e(\Tb{2})$.
However, this would be only partially correct as some transitions are observed more often, and hence, should have a higher weight.
To account for this, we weight each $h_e$ by the relative frequency of $e$.
In formula, we have:
\begin{equation}
  H( \Tb{2} ) = \sum_{e\in E} \mathbf{\pi}_{e}
h_e(\Tb{2}) = - \sum_{e\in E} \mathbf{\pi}_{e} \sum_{e'\in E} (\Tb{2} )_{ee'} \log (\Tb{2})_{ee'}
\end{equation}
where $\mathbf{\pi}_{e}$ is the relative frequency of the
transition $e$, that is given by:
\begin{equation}
    \mathbf{\pi}_{e} = \frac{|\{p : p = e \wedge p \in \mathcal{S}
\}|}{|\mathcal{S}|}
\end{equation}
where $\mathcal{S}$ is the multi-set of paths and sub-paths constructed using the empirical data.
With the above definition, we can compute the entropy growth ratio between the first- and the second order models:
\begin{equation}
  \Lambda_{H} (\Tb{2}  ) = H(\Tbt{2})  / H(\Tb{2})
\end{equation}
Note that for $\Lambda_{H} (\Tb{2} ) >1$ the diffusion process computed using the first-order network would
over-estimate the level of mixing.
If $\Lambda_{H} (\Tb{2})<0$, the opposite is true.
We obtain a $\Lambda_{H} (\Tb{2}) \sim 3.4 > 1 $.
This means that the growth of entropy coming from the first-order model is 3 times larger than expected from the second-order model.
Hence, the mixing would be overestimated by using a first-order model.
This result is in line with the qualitative findings from Fig.\ref{fig:alluvial}.

\section{City and Country Level}
\label{sec:city-and-country-lev}
We now move our analysis from the top 100 universities to the city and country level.
For this analysis, we aggregate trajectories traversing institutions located in the same cities or countries.
By not limiting the analysis to specific set of universities, we increase the number of observed trajectories and scientists.
Hence, we should expect a more reliable statistics.
At the same time, we should be aware that aggregating and projecting sequential data can distort its temporal properties as discussed previously, see also \citet{scholtes2014causality}.

We restrict our attention to $3\,740\,187$ individual scientist trajectories across 215 countries between 1990 and 2009 with papers listed in MEDLINE.
89\% of the trajectories are of length 0, meaning that most scientists do not change country.
However, this statistic also means that longer career trajectories are rare but still numerous ($411\,137$).
The most frequent trajectories of length one are between the UK and USA, Japan and the USA, and the USA and the UK.
The presence of the US in these paths stems from the fact that the US has the largest scientist population in the dataset.

When considering trajectories of length two, the most frequent ones are between (Japan, USA, Japan), (USA, UK, USA), and (UK, USA, UK).
If we consider only those trajectories that do not go through the USA,
we find that the most frequent trajectories of length two are across
(UK, Australia, UK), (France, UK, France) and (Germany, UK, Germany).
All these types of trajectories are reminiscent of the motifs of type I, i.e. $X\to Y \to X$, found at the institution level.
When looking at trajectories at the city level, we find a similar pattern (see Appendix \ref{app:city_traj}).
Therefore, we test for the existence of temporal correlations as we have done at the institution level.

By applying the multi-order modelling framework of \citet{scholtes2017kdd}, we identify $k_{\mathrm{opt}}=2$, implying that a second-order network best represents the data at the country level.
When applying the same framework at city level, we identify $k_{\mathrm{opt}}=1$.
Hence, we \emph{do not} find evidence of temporal correlations at city, but at country level.
This result has several implications.
First, even though we observe a large number of trajectories of type I at a both city and country level, these trajectories are still to be expected from a first-order network model at city, but not at country level.
Second, temporal correlations have been detected at the affiliation and country, but not at city level.
Therefore, we find that the concerns raised by~\citet{butts2009revisiting,zweig2011good,scholtes2017kdd} about the naive application of network-analytic models apply to research on academic mobility.

\section{Discussion}
\label{sec:discussion}

Scientific knowledge is shared not only through artefacts, e.g., publications, but also through scientists carrying out their research.
Thus, their mobility is thought to facilitate knowledge exchange.
``Mobility as a vehicle of knowledge exchange'' implicitly assumes that academic careers can span the globe and universities.
In this work, we have addressed this implicit assumption by testing for the existence of temporal correlations in scientific careers.
The existence of such correlations would imply the existence of constraints in academic mobility.

To perform our analysis we have reconstructed, from two large bibliographic datasets, {MAG}~\citep{sinha2015overview} and {MEDLINE}~\citep{Torvik2009, Torvik2015}, the career trajectories across the 100 top universities and across cities and countries, respectively.
We do not find evidence that temporal correlations influence career trajectories at city level.
However, we do find evidence that mobility, especially among elite institutions, is influenced by temporal correlations.

At the \emph{university level}, we find that the mobility of scientists is affected by temporal correlations.
In particular, a large part of these correlations is determined by career trajectories of type I, i.e., scientists return to their original university after two career steps.
Indeed, the KL-divergence ratio $R_I$ for these motifs is five times larger than $R_{II}$.
This result has a direct impact on the knowledge exchange as proxied by scientists mobility.
Specifically, the implied mixing of knowledge through mobility is a lot lower than one would expect, by using a standard networks model.
To quantify this statement, we have computed the entropy growth ratio $\Lambda_{H}(\Tb{2})$, that captures how much a standard network model overestimates the trajectory mixing.
We find that $\Lambda_{H}(\Tb{2})\sim 3.4$, showing that a simple network model would overestimate the mixing by a factor of 3.
A possible but rather pessimistic explanation is that a simple mental short-cut employed by hiring committees to favour ``known origins'' reinforces the path dependency and possibly hampers knowledge exchange.

At the \emph{city and country level}, we find a large number of career trajectories of type I.
However, at city level, we do not find statistically significant temporal correlations.
This result indicates that type I trajectories or returnees~\citep{Saxenian2005, Agrawal2006, Agrawal2011} are still to be expected from a standard network model.
Additionally, the use of a standard network perspective is justified when studying academic mobility at city, but not at country level.

Overall from the results presented in this work, a picture emerges that scientific careers are a lot less diverse than one would expect.
The existence of temporal correlations suggests that although among the 100 most prestigious institutions, there is substantial and continued exchange, the actual career trajectories are far more limited.
Our results are relevant both methodologically and empirically, for research on high-skill labour mobility, in general, and the analysis of scientists' career trajectories, in particular.

\section{Declarations}
\subsection{Author's contributions}
    GV, LV and FS have conceived the study.
    GV and LV have performed the analysis.
    GV, LV and FS have written the manuscript.

\subsection{Acknowledgements}
  The authors thank Dr. Giona Casiraghi for his useful suggestions on how to present the analysis for the entropy growth ratio.

\subsection{Availability of data and materials}
The raw XML data on all MEDLINE articles is available for download from the NIH at
\begin{itemize}
\item \url{https://www.nlm.nih.gov/databases/download/pubmed_medline.html}
\item \url{ftp://ftp.ncbi.nlm.nih.gov/pubmed/baseline}
\end{itemize}

The disambiguation of authors (Authority)~\citep{Torvik2009} and affiliations (MapAffil)~\citep{Torvik2015} has been obtained from
\begin{itemize}
\item \url{http://abel.lis.illinois.edu/downloads.html}
\end{itemize}
Access to this resource can be requested for free from the maintainers through the online form on the same page.
Note that due to an agreement with the providers of Authority and MapAffil, these datasets may only be shared by requesting access through the previously mentioned online form.

The MAG data~\citep{sinha2015overview} was downloaded from \url{https://www.kdd.org/kdd-cup/view/kdd-cup-2016/Data}.
Unfortunately, the data has now been removed for download.
Hence, the MAG data used and analysed are available from the corresponding author on reasonable request.

\bibliographystyle{sg-bibstyle-nourl} 
\bibliography{bibliography}  
\appendix
\include{appendix}

\end{document}

%% file: appendix.tex
\section{Distributions of motifs}\label{app:motifs}
\subsection{Computing the distribution probabilities of motifs}\label{app:comp-motifs}
Let us define the \textit{multi-set} $\mathcal{S}$ to contain all the trajectories present in the data.
For example, given a trajectory $i = \{A\to B\to A\to B\}$ appearing $n_i$ times in the data, then $i$ appears $n_i$ times in $\mathcal{S}$.
For each trajectory $i\in\mathcal{S}$, we also define the \textit{multi-se}t $\mathcal{S}_k(i)$ to contain all the sub-trajectories of $i$ with length $k$.
For example, given a trajectory $i = \{A\to B\to A\to B\}$, the \textit{multi-set} $\mathcal{S}_1(i)=\{A\to B, A\to B, B\to A\}$.
Note that the sub-trajectory $A\to B$ appears twice in $i$, i.e., $m(A\to B, i) = 2$, while $B\to A$ appears only once, i.e., $m(B\to A, i) = 1$.
In general, we have that $|\mathcal{S}_k(i)| = \max(l_i - k + 1, 0)$ where $l_i$ is the length of trajectory $i$.
In other words, a trajectories of length $l_i$ can be split into $l_i - k + 1$ sub-string of length $k$ if $k \leq l_i$. 
This can also be expressed as $\sum_{p\in\tilde{\mathcal{S}}_k(i)} m(p,i) = \max(l_i - k + 1, 0)$ where $\tilde{\mathcal{S}}_k(i)$ is the \textit{set} of sub-trajectories of length $k$ extracted from $i$.

To compute the probability to observe motifs of type I, i.e., $X \to Y \to X$ or type II, i.e., $X \to Y \to Z$, we have note that these are all the possible (sub-)trajectories of length two. 
We call $\mathcal{S}_k$ the \textit{multi-set} containing all the sub-trajectory of length $k$.
This means that $\mathcal{S}_2$ contains all the motifs of type I and type II.
Then, we define the \textbf{probability to observe a motif} $p \in \mathcal{S}_2$ 

\begin{equation}\label{eq:emp_motifs_dist}
    P ^{ (\textrm{emp}) } (p) =  \frac{1}{|\mathcal{S}|}\sum_{i\in\mathcal{S}} q(p|i) =  \frac{1}{|\mathcal{S}|}\sum_{i\in\mathcal{S}} \frac{m(p,i)}{\max(l_i - 1, 1)}
\end{equation}

Note that $P ^{ (\textrm{emp}) }$ is is always equal or greater than zero and $\sum_{p \in \tilde{\mathcal{S}_2}}P ^{ (\textrm{emp}) }(p)  + |\mathcal{S}_1|/|\mathcal{S}|  =1 $ where $\tilde{\mathcal{S}_2}$ is the \textit{set} containing the sub-trajectory of length two.
This last relation follows from the fact that any trajectory $i\in\mathcal{S}$ can be spitted in $l_i - 1$ sub-trajectories of length $2$ when $l_i \geq 2$.
Hence, 
\begin{equation}\label{eq:motfis_cases}
    \sum_{p \in \tilde{\mathcal{S}_2}} q(p|i) =
    \begin{cases}
        1  & \forall i \,: \,l_i \geq 2 \\
        0  & \forall i \,: \,\textrm{otherwise}
    \end{cases}
\end{equation}

Using Eq.\ref{eq:motfis_cases}, we can show that
\begin{equation}
\sum_{p \in \tilde{\mathcal{S}_2}}P ^{ (\textrm{emp}) }(p) = 
    \frac{1}{|\mathcal{S}|}\sum_{i\in\tilde{\mathcal{S}}} n_i \sum_{p \in \tilde{\mathcal{S}}_2}q(p|i) =
    \frac{1}{|\mathcal{S}|}\sum_{k=2} |\mathcal{S}_k|
    \Longrightarrow 
    \sum_{p \in \tilde{\mathcal{S}}_2}P ^{ (\textrm{emp}) }(p)  + \frac{|\mathcal{S}_1|}{|\mathcal{S}|} =1
\end{equation}

 For computing the probability distribution in the first- and second-order network, i.e., $P^{(1)}$ and $P^{(2)}$, we rely on the \texttt{Pathpy} implementation.
 Precisely, we use the function \texttt{path\_likelihood} that returns the probability to observe a path/transition.
 For more details, see~\citep{scholtes2017kdd,pathpy}.

 \subsection{Odds ratio}

We split the set $\tilde{\mathcal{S}_2}$ in two disjoint sets $\tilde{\mathcal{S}}^{I}_2$ and $\tilde{\mathcal{S}}^{II}_2$,respectively containing motifs of type I and type II.
The empirical odds ratio to observe motifs of type I compared to motifs of type II is given by

\begin{equation}
    OR_{I,II} = \frac{\sum_{p \in \tilde{\mathcal{S}}^{I}_2}P ^{ (\textrm{emp}) }(p)}{\sum_{p \in \tilde{\mathcal{S}}^{II}_2}P ^{ (\textrm{emp}) }(p)} 
\end{equation}

 and we find from our data that $OR_{I,II} \sim 1.7$. This means that motifs of type I are almost twice as frequent as motifs of type II.
When computing the odds ratio $OR_{I,II}$ using $P^{(1)}$ we get $\sim 0.14$ that is of an order magnitude different compared to the empirical one.
While, the the odds ratio $OR_{I,II}$ coming from $P^{(2)}$ is $\sim 1.4$ that is quite close to the empirical one.

\subsection{Computing and visualizing the Kullback-Leibler divergence}

\begin{figure}
    \centering
    \includegraphics[width = 0.45\textwidth]{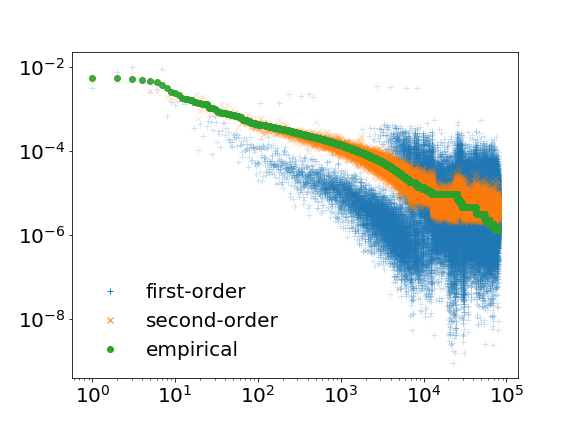}
    \caption{Distribution probabilities for all the motifs of type I and II.
    we plot $P ^{ (\textrm{emp}) }$, $P^{(1)}$, and $P^{(2)}$ respectively in green dots, blue $+$ and orange $\times$.
The order on the x-axis is created by sorting the motifs from the most to least probable with respect to the empirical data.
On the y-axis, we report their respective probabilities.}
\label{fig:kl-vis-all}
\end{figure}

Since we use the $P ^{ (\textrm{emp}) }$, $P^{(1)}$, $P^{(2)}$ to compute the Kullback-Leibler (KL) divergence for motifs of length 2, we normalize this probability to one.
In other words, for $P ^{ (\textrm{emp}) }(p)$, we write 
\begin{equation}
P ^{ (\textrm{emp}) }(p) \to P ^{ (\textrm{emp}) }(p) / \sum_{p \in \tilde{\mathcal{S}}_2}P ^{ (\textrm{emp}) }(p)
\end{equation}
and we renormalise in similar way  $P^{(1)}$ and $P^{(2)}$.
The KL-divergence between $P ^{ (\textrm{emp}) }$ and $P^{(1)}$ is 1.51, while between $P ^{ (\textrm{emp}) }$ and $P^{(2)}$ is 0.10.
Hence, the first-order model has KL-divergence more than 10 times larger compared to the second-order model.
In Fig.~\ref{fig:kl-vis-all}, we plot $P ^{ (\textrm{emp}) }$, $P^{(1)}$, and $P^{(2)}$ respectively in green dots, blue $+$ and orange $\times$.
The order on the x-axis is created by sorting the motifs from the most to least probable with respect to the empirical data.
We see that the first-order network consistently underestimates motifs that are more probable.
Instead, the second-order network produces probabilities quite close to the empirical ones.

\begin{figure}
    \centering
\includegraphics[width = 0.45\textwidth]{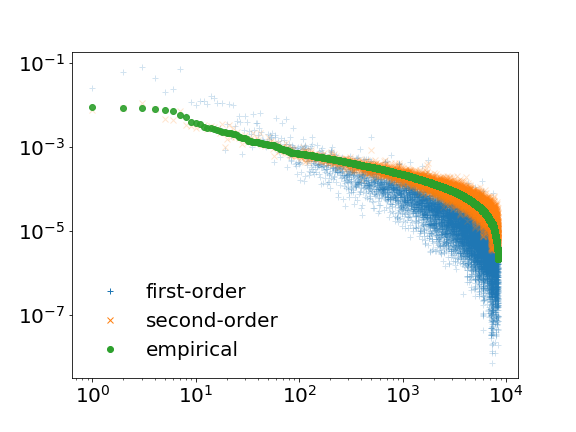}
\includegraphics[width = 0.45\textwidth]{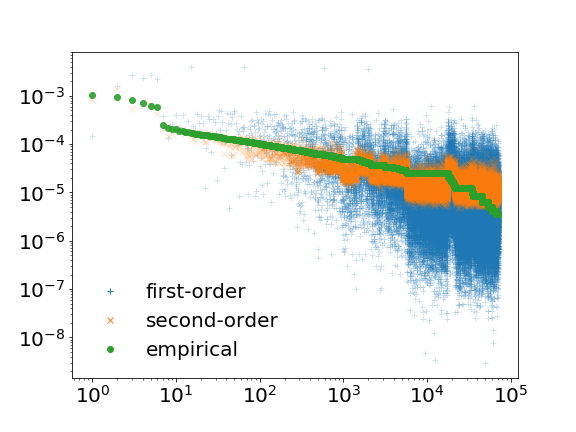}
\caption{Distribution probabilities for all the motifs of type I (a) and II (b).
we plot $P ^{ (\textrm{emp}) }$, $P^{(1)}$, and $P^{(2)}$ respectively in green dots, blue $+$ and orange $\times$.
The order on the x-axis is created by sorting the motifs from the most to least probable with respect to the empirical data.
On the y-axis, we report their respective probabilities.}
\label{fig:kl-vis}
\end{figure}

To compute the Kullback-Leibler divergence only for motifs of type I (type II), we have to renormalise  $P ^{ (\textrm{emp}) }$, $P^{(1)}$, and $P^{(2)}$.
In other words, $P ^{ (\textrm{emp}) }(p) \to P ^{ (\textrm{emp}) }(p) / \sum_{p \in \tilde{\mathcal{S}}^{I}_2}P^{ (\textrm{emp}) }(p)$, and similarly for  $P^{(1)}$ and $P^{(2)}$.
In Fig.~\ref{fig:kl-vis}(a) and (b), we compare the probability distribution for the motifs of type I and type II, respectively.
Again, we have $P^{ (\textrm{emp}) }$, $P^{(1)}$, and $P^{(2)}$ respectively in green dots, blue $+$ and orange $\times$.
The order on the x-axis is created by sorting the motifs from the most to least probable with respect to the empirical data.
From Fig.~\ref{fig:kl-vis}(a) that depict motifs of type I, it is evident that the first order network underestimates the probabilities of these motifs.
While, in Fig.~\ref{fig:kl-vis}(b), the situation is reversed.

\section{Trajectories at city level}
\label{app:city_traj}
In MEDLINE, we have $3\,740\,187$ individual scientist trajectories across 12 980 cities between 1990 and 2009..
Among these, $884\,251$ trajectories have length 1 or higher.
Specifically, 11 \% of all the trajectories are of length 1, meaning that we observe half of the scientists changing city only once. 
While, the 12 \% of the trajectories are longer (i.e., $455\,127$).
The most frequent trajectories of length one are between 
 Boston (MA, USA) and Cambridge (MA, USA),
 London (UK) and the Oxfordshire (UK), 
 and Tokyo (Japan) and Kanagawa (Japan).

While the most frequent trajectories of length 2 are  between 
Boston (MA, USA), Cambridge (MA, USA) and Boston (MA, USA),
Stanford, (CA, USA), Palo Alto, (CA, USA), and Stanford, (CA, USA)\footnote{Note that the fact that we can distinguish between locations such as Stanford and Paolo Alto only thanks to the fine grain resolution of MapAffil~\citep{Torvik2015}.
Indeed, by manually checking authors' affiliation that MapAffil placing in Palo Alto, we find that many companies such Xerox, Hewlett-Packard, Lockheed Martin, and many biotech companies have laboratories in Paolo Alto.},
London (UK) Oxfordshire (UK), London (UK),
Tokyo (Japan), Kanagawa (Japan), Tokyo (Japan).

\section{Trajectories at global affiliation level (MAG)}
\label{app:all_aff_traj}
In the MAG, we have over $14$ million disambiguated scientist among more than $19 000$ affiliations.
Among these, $2\,591\,784$ trajectories have length 1 or higher between $18\,522$.
Specifically, $1\,196\,158$ are of length 1, meaning that we observe half of the scientists changing affiliation only once. 
While, $1\,395\,626$ of the trajectories are longer (e.g., $614\,776$ have length two).
The most frequent trajectories of length one are between 
University of California and University of California, Berkeley
University of California and university of California Los Angeles, and
University of California and Davis, University of California.
Note that University of California is a university system composed of 10 different campuses 
and is not an unambiguously defined location or affiliation.
When considering longer trajectories, we find similar type of trajectories trough ambiguously defined research institutions.
For this reasons, we do not use the MAG data to analyze scientists' career trajectories at global level and use MEDLINE instead.